\documentclass[letterpaper, final, twocolumn, showpacs, pre,
tightenlines, lengthcheck, raggedbottom, superscriptaddress]{revtex4}

\pdfoutput=1

\usepackage{mathptmx}
\usepackage{amsmath}
\usepackage{graphicx}

\usepackage{hyperref}

\newcommand{\x}{\mathbf{x}}
\renewcommand{\v}{\mathbf{v}}
\newcommand{\DMZ}{D_\mathrm{MZ}}
\newcommand{\DoneKK}{D^{(1)}_\mathrm{KK}}
\newcommand{\Done}{D_\mathrm{1SMA}}
\newcommand{\DtwoKK}{D^{(2)}_\mathrm{KK}}
\newcommand{\Dtwo}{D_\mathrm{2SMA}}
\newcommand{\pf}{P_\mathrm{f}}
\newcommand{\pb}{P_\mathrm{b}}
\newcommand{\pr}{P_\mathrm{r}}
\renewcommand{\pl}{P_\mathrm{l}}
\newcommand{\ps}{P_\mathrm{s}}
\newcommand{\pff}{P_\mathrm{ff}}
\newcommand{\pfb}{P_\mathrm{fb}}
\newcommand{\pfl}{P_\mathrm{fl}}
\newcommand{\pfr}{P_\mathrm{fr}}
\newcommand{\pbf}{P_\mathrm{bf}}
\newcommand{\pbb}{P_\mathrm{bb}}
\newcommand{\pbr}{P_\mathrm{br}}
\newcommand{\pbl}{P_\mathrm{bl}}
\newcommand{\prf}{P_\mathrm{rf}}
\renewcommand{\prb}{P_\mathrm{rb}}
\renewcommand{\prl}{P_\mathrm{rl}}
\newcommand{\prr}{P_\mathrm{rr}}
\newcommand{\plf}{P_\mathrm{lf}}
\newcommand{\plb}{P_\mathrm{lb}}
\newcommand{\plr}{P_\mathrm{lr}}
\newcommand{\pll}{P_\mathrm{ll}}
\newcommand{\psf}{P_\mathrm{sf}}
\newcommand{\psb}{P_\mathrm{sb}}

\newcommand{\pss}{P_\mathrm{ss}}
\newcommand{\mean}[1]{\left \langle #1 \right \rangle}

\newcommand{\ii}{i}

\begin{document}

\title{Persistence effects in deterministic diffusion}
\author{Thomas Gilbert}
\email{thomas.gilbert@ulb.ac.be}
\affiliation{Center for Nonlinear Phenomena and Complex Systems,
  Universit\'e Libre  de Bruxelles, C.~P.~231, Campus Plaine, B-1050
  Brussels, Belgium}

\author{David P.~Sanders}
\email{dps@fciencias.unam.mx}
\homepage{http://sistemas.fciencias.unam.mx/~dsanders}
\affiliation{Departamento de F\'isica, Facultad de Ciencias, Universidad
Nacional Aut\'onoma de M\'exico,  Ciudad Universitaria,
04510 M\'exico D.F.,
 Mexico}

\begin{abstract}
In systems which exhibit deterministic diffusion, the gross parameter
dependence of the diffusion coefficient can often be understood in terms of
random walk models. Provided the decay of correlations is fast
enough,
one can ignore memory effects and approximate the diffusion coefficient
according to dimensional arguments. By successively including the effects
of one and two 
steps of memory on this approximation, we examine the effects of 
``persistence'' on
the diffusion coefficients of extended two-dimensional billiard tables and
show how to properly account for these effects, using walks in
which a particle undergoes jumps in different directions with probabilities
that depend on where they came from.
\end{abstract}


\maketitle

\section{Introduction}

Diffusion is a fundamental macroscopic phenomenon in physical systems,
which, for instance, characterizes the spreading of tracer particles in a
solvent. At a mescoscopic scale, it can be traced to the cumulative effect
of many ``random'' displacements, as in Brownian motion \cite{Cha43}. At the
underlying microscopic scale, however, the dynamics of a system are
deterministic. Deterministic diffusion concerns the study of microscopic
models whose deterministic dynamics also exhibit diffusive behavior at
a macroscopic scale \cite{Gas98,Dor99,Kla07}. 

A particularly appealing, physically motivated model which does exhibit this
phenomenon is the periodic Lorentz gas \cite{BS80}. Here, independent point
particles in free motion undergo elastic collisions with fixed hard disks
in a periodic array. The diffusive motion can then be considered to be a
result of the chaotic nature of the microscopic dynamics, 
according to which nearby initial conditions separate exponentially fast
due to the convex nature of the obstacles. Thus a cloud of
(non-interacting) particles in this Lorentz gas spreads out over time in a
way similar to that of solutions of the diffusion equation, 
\begin{equation}
 \mean{[\x(t) - \x(0)]^2} \sim 4 D t,
\end{equation}
where $\x(t)$ denotes the position of a tracer at time $t$, with initial
position $\x(0)$, and the mean squared displacement is computed as an
average $\mean{\cdot}$ over many
realizations of this process. 
The diffusion coefficient, $D$, is a constant
which depends on the geometrical parameters of the system, i.e., the
underlying microscopic dynamics.

The diffusion coefficient summarizes the macroscopic behavior of the
system while capturing the microscopic properties of the dynamics that lead
to it. A central question in deterministic diffusion is to understand how this
dependence on the geometrical parameters comes about. This has
been addressed in particular by Machta and Zwanzig \cite{MZ83}, who showed
that in the limit where the obstacles are close together, the motion
reduces to a stochastic Bernoulli-type hopping process---a \emph{random
walk}---between
``traps''. By calculating the diffusion coefficient of this random walk,
they were able to obtain a reasonable agreement with the
numerically-measured value of the diffusion coefficient. 

The approach of Machta and Zwanzig was extended heuristically by Klages and
Dellago \cite{KD00}, by including important physical effects not taken
into account in the simple random-walk picture, namely a possibly
non-isotropic probability of changing directions, and of crossing two traps
at once. Klages and Korabel \cite{KK02} then provided an alternate 
approach, in which they employed a Green-Kubo expansion of the diffusion
coefficient to obtain a series of increasingly accurate approximations,
based on numerically-calculated multi-step transition probabilities. In one
particular Lorentz gas model, they showed that their results are in good
agreement with this expansion, see also \cite{Kla07}. Nonetheless, as we
emphasize below, the physical motivation, and indeed the physical meaning,
of this approach, are not clear. 

The purpose of this paper is to show that in fact the correct expansion
beyond the Machta-Zwanzig approximation is to incorporate this type of
correction in the framework of \emph{persistent random walks}. In other
words, to be consistent, memory effects of a given length, whether one or
several steps, must be accounted for through their contribution at all
orders in the Green-Kubo formula relating the diffusion coefficient to the
velocity auto-correlations. This is physically strongly motivated, and
provides the correct way of incorporating correlation effects, in
principle, of any finite order. 

In the usual periodic Lorentz gas on a triangular lattice considered in
\cite{KK02}, the model is sufficiently isotropic that higher order
contributions
are small and can be safely neglected. However, when the correlation
effects are very strong, this is no longer the case. We introduce a
billiard model with
this property and show that, whereas a Green-Kubo--type expansion fails
except very close to the Machta-Zwanzig limiting case, the approximation
based on a first-order persistent random walk is in reasonable agreement
with the data. By further considering memory effects up to two successive
steps, we find that the agreement between the numerically-measured
diffusion coefficient of the billiard table and that of the second-order
persistent random walk extends to an appreciably larger range of 
parameters. 

\section{Periodic Lorentz gas on a triangular lattice}

\begin{figure}[t]
  \centering
  \includegraphics[width=0.45\textwidth]{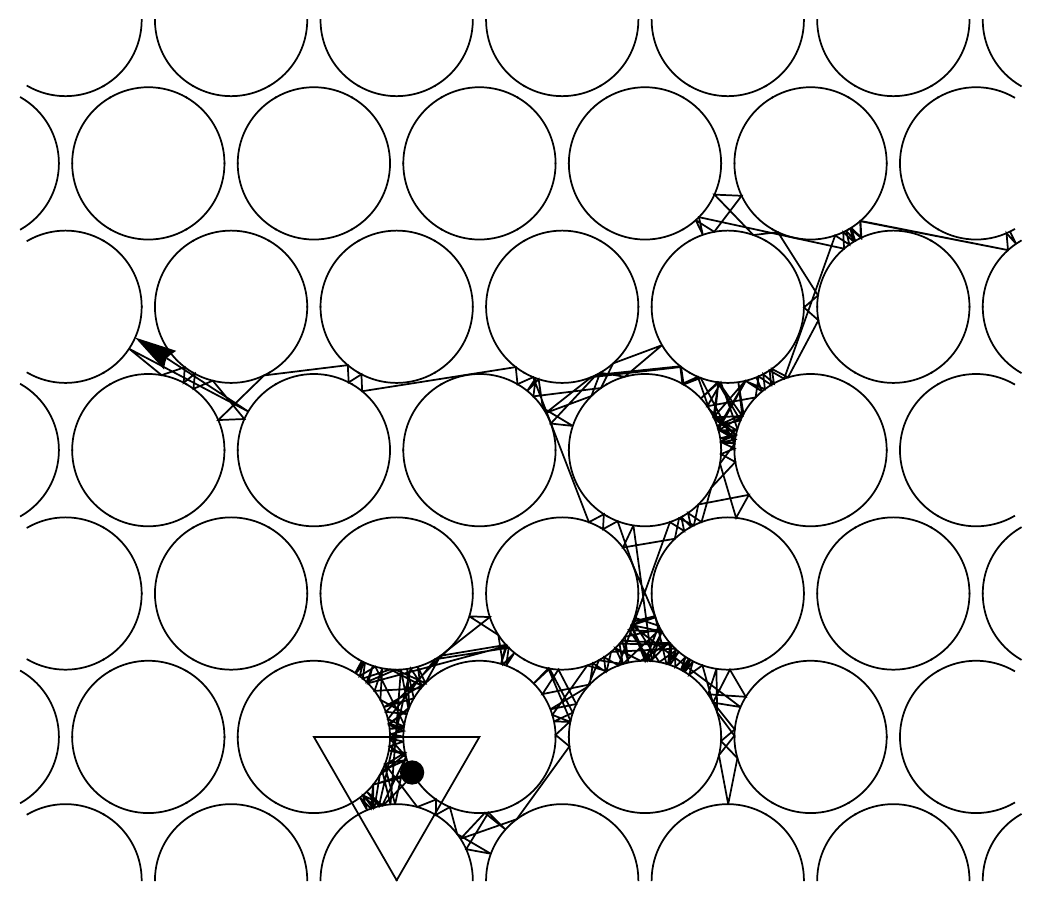}
  \caption{Periodic Lorentz gas on a triangular lattice. A typical
    trajectory is shown, which starts at the upper right disk in the
    initial cell---marked by the highlighted triangle---and moves across
  the table, performing a diffusive motion.}
  \label{fig.table-hc}
\end{figure}

Consider the periodic Lorentz gas on a triangular lattice, shown 
in Fig.~\ref{fig.table-hc}. The centers of three nearby disks are
identified as the vertices of equilateral triangles which, in our notation,
will be taken to be of unit side length. Denoting by $\rho$ the radius of the
disks, we let $\delta \equiv 1 - 2 \rho$ denote the spacing between disks. When
$\delta = 0$, the triangles form closed traps from which the tracer
particles cannot escape. If, however, $0 < \delta \ll 1$, then we expect the
particles to remain inside the cell for a long time before they can escape
to another cell. This argument can be made precise using ergodic
theory \cite{MZ83}. Given a tracer of unit velocity, the mean trapping time
$\tau$ is given in
terms of the ratio of the cell area to the lengths of the holes by \cite{CM06}
\begin{equation}
  \tau
 = \pi \frac{\sqrt{3}/4 - \pi \rho^2/2}{3\delta}
  = \pi \frac{\sqrt{3}/4 - \pi (1-\delta)^2/8}{3\delta}\,.
  \label{tauhc}
\end{equation}
Assuming that this time is longer than the typical decay of correlations
\footnote{The exponential decay of correlations has been proved in the
  periodic Lorentz gas, see N.~Chernov and L.-S.~Young, \emph{Decay of
    correlations for Lorentz gases and hard balls}, in: \emph{Hard Ball
    Systems and the Lorentz Gas}, ed.\ by D.~Szasz, Encyclopaedia of
  Mathematical Sciences \textbf{101}, 89 (Springer, 2000).}, 
Machta and Zwanzig \cite{MZ83} argued that the diffusion coefficient of the
Lorentz gas can be approximated by a memory-less random walk (called 
the ``short-memory approximation'' in \cite{Z83}),
\begin{equation}
  \DMZ = \frac{\ell^2}{4\tau
}\,,
  \label{DMZ}
\end{equation}
where, in our notation, the lattice spacing $\ell = 1/\sqrt{3}$.

Taking further account of memory effects, Klages \& Korabel \cite{KK02}
noted that the Machta-Zwanzig approximation (\ref{DMZ}) is the zeroth order
expansion of a series given by the Green-Kubo formula for the diffusion 
coefficient:
\begin{equation}
  D = \DMZ \left( 1 + 2 \sum_{k=1}^\infty \mean{\v_0
      \cdot \v_k} \right), 
  \label{dcoef}
\end{equation}
where $\v_k$ is the jump vector (``velocity'') between traps on the $k$th
step, and $\mean{\v_0 \cdot \v_k}$ are \emph{auto-correlations} of the
velocity at steps $0$ and $k$.

The approach taken in \cite{KK02} was to truncate the expression
\eqref{dcoef}  at a finite value of $k$, assuming that all
higher-order correlations are $0$. They showed that by numerically
calculating the terms appearing in this equation, the results evaluated by
this truncation converged to the numerically-obtained diffusion
coefficient. 

However, this ad hoc truncation has no physical meaning: if $\mean{\v_0 \cdot
  \v_1} \neq 0$,  it is not true that higher-order correlations $\mean{\v_0
  \cdot \v_k}$ vanish. Rather, assuming that the process has memory of
the previous step alone---which we will refer to as the single-step memory
approximation---one must compute the correlations $\mean{\v_0 \cdot \v_k}$
in a consistent way, and evaluate the diffusion coefficient (\ref{dcoef}) by
taking all the $k$'s into consideration.

\subsection{Single-step memory approximation}

The process for which the motion of a particle at step $k+1$ depends
explicitly on the state at step $k$ is known as a \emph{persistent} or
\emph{correlated} random walk \cite{WR83,Hu95,HK87,We94}. The technique for
studying such walks is well developed, and consists of treating it
as a random walk with internal states, which, in this case,
describe the direction with which the walker arrives at a site. 

Considering a persistent random walk on a honeycomb lattice, we denote by
$\pb$ the conditional probability to return in the direction
opposite to the current one, $\pr$ to turn right and $\pl$ to turn 
left. In terms of these quantities, the velocity auto-correlation is found
to be \cite{GS09}
\begin{eqnarray}
  \mean{\v_0 \cdot \v_k}
  &=& \frac{(-1)^k}{2} \Bigg\{
    \left[\pb - \frac{\pr + \pl}{2} - \ii\frac{\sqrt{3}}{2}
      (\pr - \pl)\right]^k  
    \nonumber\\
    && +      
    \left[\pb - \frac{\pr + \pl}{2} + \ii\frac{\sqrt{3}}{2}
      (\pr - \pl)\right]^k  
  \Bigg\}\,,
\label{v0vkhc}
\end{eqnarray}
where $\ii = \sqrt{-1}$. In the case of a symmetric walk for which $\pr
= \pl \equiv \ps = (1 - \pb)/2$, the diffusion coefficient (\ref{dcoef}) is
\begin{equation}
  \Done = \DMZ\frac{3(1 - \pb)} {1 + 3 \pb}\,.
  \label{D1SMAhc}
\end{equation}
In comparison, the first order approximation made in \cite{KK02} is to write
\begin{equation}
  \DoneKK = 1 + 2\mean{\v_0 \cdot \v_1} = \DMZ (2 - 3 \pb)\,,
  \label{DKK1hc}
\end{equation}
which corresponds to the first-order approximation of Eq.~(\ref{D1SMAhc})
when expanding $\Done$ about the isotropic process for which $\pb = 1/3$. 

\subsection{Two-step memory approximation}

For a persistent process with two-step
memory approximation, there are nine conditional transition
probabilities, which we denote by $\pbb$, $\pbr$, $\pbl$, $\prb$, $\prr$,
$\prl$, $\plb$, $\plr$ and $\pll$, where, for instance, $\pbr$ denotes the
probability that the tracer first moves backwards and then turns right, and
similarly for the other symbols. Although the
corresponding Markov chain involves a $9\times9$ 
stochastic matrix, there are symmetries in the system which can be
exploited to reduce the computations to $3\times3$ matrices involving the
transition probabilities listed above. The computation of the velocity
auto-correlation then yields \cite{GS09} 
\begin{widetext}
\begin{equation}  
  \langle \v_{0}\cdot\v_k\rangle
  =  (-1)^{k} \frac{3}{2}
  \Big(1\enspace 1\enspace 1\Big)
  \left[
    \left(
      \begin{array}{c@{\enspace}c@{\enspace}c}
        \pbb& \plb& \prb \\
        \phi \pbl& \phi \pll& \phi \prl \\
        \phi^2 \pbr& \phi^2 \plr& \phi^2 \prr
      \end{array}
    \right)^{k-1}
    \left(
      \begin{array}{c@{\enspace}c@{\enspace}c}
        1 & 0 & 0 \\
        0 & \phi & 0 \\
        0 & 0 & \phi^2
      \end{array}
    \right)
    +
    \left(
      \begin{array}{c@{\enspace}c@{\enspace}c}
        \pbb& \plb& \prb \\
        \phi^2 \pbl& \phi^2 \pll& \phi^2 \prl \\
        \phi \pbr& \phi \plr& \phi \prr
      \end{array}
    \right)^{k-1}
    \left(
      \begin{array}{c@{\enspace}c@{\enspace}c}
        1 & 0 & 0 \\
        0 & \phi^2 & 0 \\
        0 & 0 & \phi
      \end{array}
    \right)\right]
  \left(
    \begin{array}{c}
      p_1\\
      p_2\\
      p_3
    \end{array}
  \right)\,,
\end{equation}
where $\phi = \exp(2\ii\pi/3)$ and $(p_1,p_2,p_3)$ are the first three
components of the stationary distribution of the Markov chain, which, for a
left-right symmetric process for which $\prr = \pll \equiv \pss$, $\prb =
\plb \equiv \psb$ and $\prl = \plr = 1 - \pss - \psb$,
read 
\begin{equation}
  p_1 = \frac{\psb}{3 - 3 \pbb + 3 \psb},\quad
  p_2 = p_3 = \frac{1 - \pbb}{6 - 6 \pbb +  6 \psb}.
  \label{invphc}
\end{equation}
The corresponding diffusion coefficient (\ref{dcoef}) is 
\begin{equation}
  \Dtwo = \DMZ
 \frac{3 (1 - \pbb) (1 + \pbb - \psb) (2 - \psb - 2 \pss)}
  {(1 - \pbb + \psb) 
    [\psb (7 + \pbb - 8 \pss) + 2 (1 + \pbb) \pss -4 \psb^2 ]}\,.
  \label{D2SMAhc}
\end{equation}
\end{widetext}
This compares to the second-order approximation following the truncation
scheme in \cite{KK02},
\begin{eqnarray}
  \DtwoKK &=& \DoneKK + 2 \mean{\v_0\cdot\v_2}\,,\nonumber\\ 
  &=& \DMZ \frac{5 - 5 \pbb - 7 \psb + 9 \pbb \psb}
  {2 - 2 \pbb + 2 \psb}\,.
  \label{DKK2hc}
\end{eqnarray}
We note that this expression is actually different from that given in
\cite{KK02}, where the stationary distribution (\ref{invphc}) was erroneously
assumed to be uniform.

\subsection{Numerical results}

\begin{figure}[htb]
  \centering
  \includegraphics[width=0.45\textwidth]{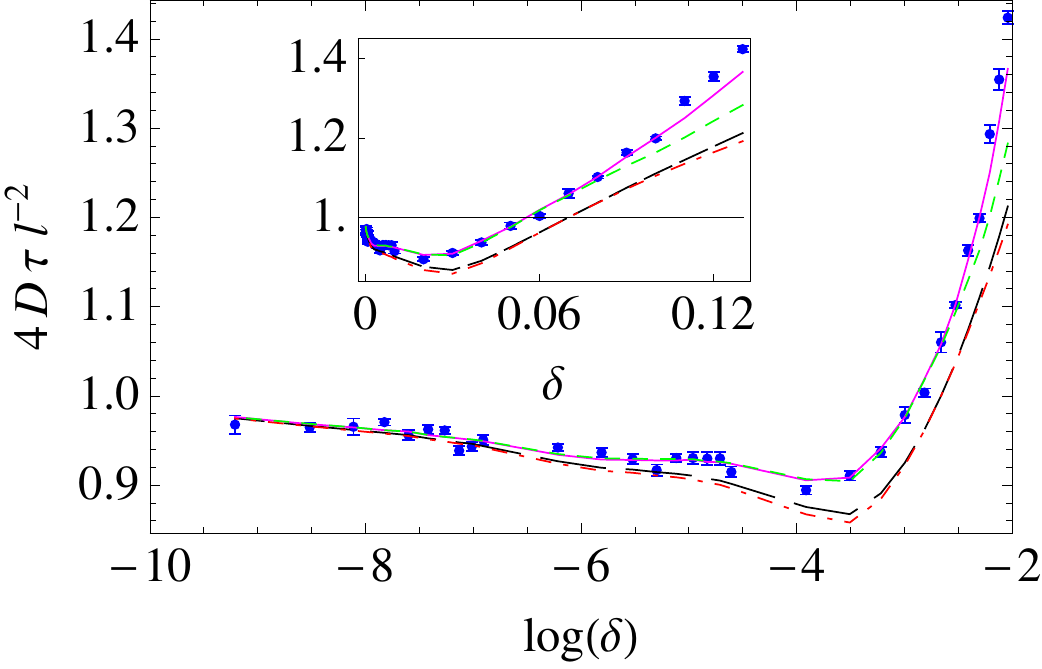}
  \caption{Numerical computation of the diffusion coefficient of the
    periodic Lorentz gas on a triangular lattice, here divided by the
    dimensional factor, Eq. (\ref{DMZ}). The lines correspond to the
    different approximate results discussed above: (long-dashed, black
    line) single-step memory approximation (\ref{D1SMAhc}); (dot-dashed,
    red line) first-order truncation (\ref{DKK1hc}); (solid, magenta line)
    two-step memory  approximation (\ref{D2SMAhc}); (dashed green)
    second-order truncation (\ref{DKK2hc}).}  
  \label{fig.dcoeff-hc}
\end{figure}

The transition probabilities of the single- and two-step memory approximation
random walks can be computed for the Lorentz gas by estimating the relative
frequencies of the corresponding events and taking into consideration the
left-right symmetry of these transitions. Plugging their values into
Eqs. (\ref{D1SMAhc}) and (\ref{D2SMAhc}), we obtain the corresponding
coefficients and compare them to the diffusion
coefficient of the billiard calculated from direct numerical simulations of the
billiard dynamics. These results are shown in
Fig.~\ref{fig.dcoeff-hc}, including the results of the truncations
(\ref{DKK1hc}) and (\ref{DKK2hc}). 

In the limit $\delta\to0$, we see that
the Machta-Zwanzig approximation (\ref{DMZ}) is recovered. Looking at a
broader range of parameter values, whereas the
single-step approximation yields a good estimate of the actual diffusion
coefficient of the Lorentz 
gas only for values of $\delta \lesssim 10^{-3}$, the extent of the range of
validity of the two-step approximation is much larger, $\delta \lesssim
0.1$.

\begin{figure}[htb]
  \centering
  \includegraphics[width=0.45\textwidth]{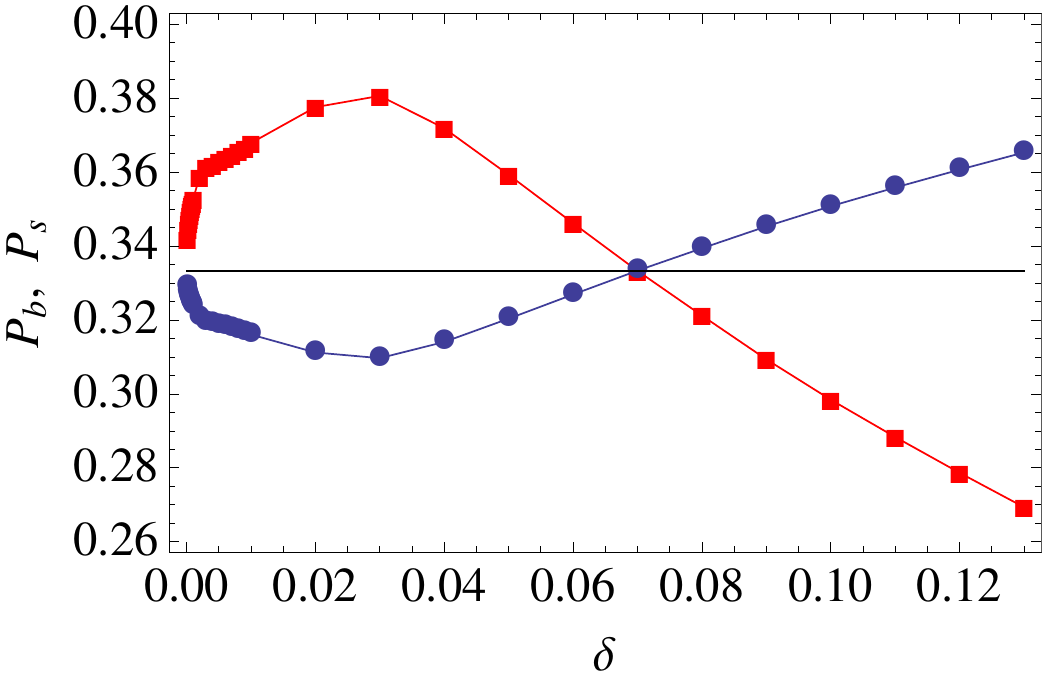}
  \caption{(Color online) Numerical computation of the transition
    probabilities $\pb$ (squares) and $\ps$ (circles) of the
    single-step memory approximation associated to the periodic Lorentz gas on
    a triangular lattice. Similar results were reported in \cite{KD00}.}
  \label{fig.corjump-hc}
\end{figure}

As seen in the figure, the successive approximations (\ref{D1SMAhc}) and
(\ref{D2SMAhc}) are slightly better than their respective counterparts
(\ref{DKK1hc}) and (\ref{DKK2hc}). The differences between the results of
Eqs. (\ref{D1SMAhc}) and (\ref{DKK1hc}) on the one hand, and
Eqs. (\ref{D2SMAhc}) and (\ref{DKK2hc}) on the other hand are, however,
quite small and not everywhere easy to appreciate. The
reason is that the transition probabilities are nearly isotropic 
for all values of $\delta$ in the range of allowed values, viz. 
$\sqrt{3}/4 < \delta < 1/2$ (the lower bound corresponds to the 
finite-horizon condition). This fact, which we illustrate in
Fig.~\ref{fig.corjump-hc} for the transition probabilities of the single-step
memory approximation, explains the validity of the Klages--Korabel truncation
scheme in this case.

\section{Billiard table on a square lattice}

\begin{figure}[htb]
  \centering
  \includegraphics[angle=270,width=0.45\textwidth]{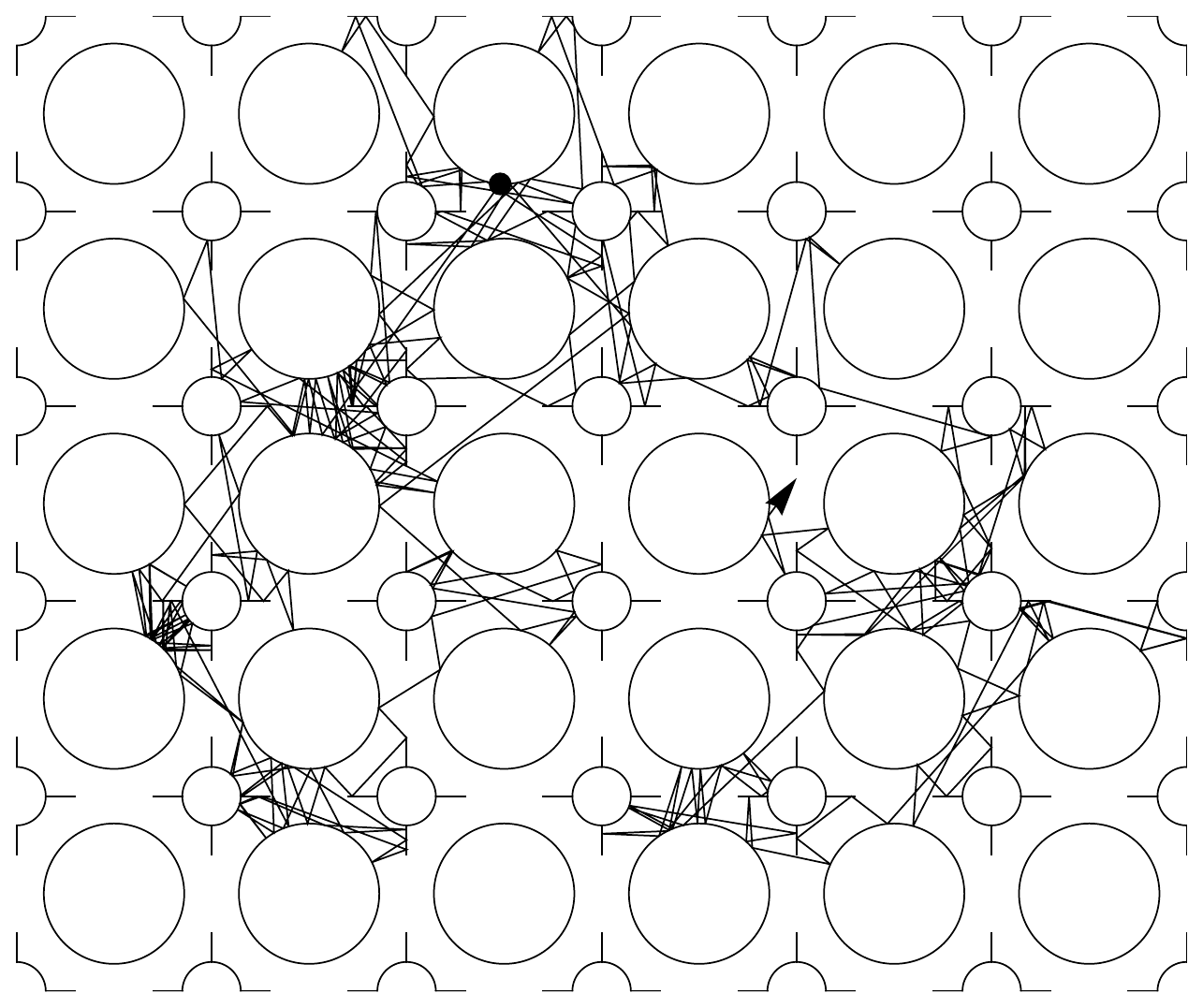}
  \caption{Periodic Lorentz table on a square lattice. A typical
    trajectory is shown, with the initial position marked by a thick
    dot. We take the inner and outer radii to be respectively
    $\rho_\mathrm{i} = 0.36 \ell$ and $\rho_\mathrm{o} = 0.15 \ell$ and
    vary only the size of the gaps.}  
  \label{fig.table-sq}
\end{figure}

In order to better emphasize the validity of our approximation scheme over
simpler truncation methods, we turn to a class of billiard tables such as
shown in Fig.~\ref{fig.table-sq}, which alternates small and large disks on
a square lattice. By inserting rigid horizontal and vertical barriers
between the small disks, with gaps of size $\delta$ in their centers, we
introduce a control parameter of relevance to the dynamical properties of the
model which is independent of the other system parameters. 
However large the
gaps, the trajectories in these billiards can be reconstructed from the
trajectories
of the same  billiard with a single unit cell on the torus.

The geometry of the lattice is such that the distance between neighboring
disks is equal to the lattice spacing, which we take to be unity, $\ell =
1$. Given the radii $\rho_\mathrm{i}$ of the large disks and
$\rho_\mathrm{o}$ of the small disks, the trapping time is  
\begin{equation}
  \tau
= \pi\frac{1 - \pi (\rho_\mathrm{i}^2 +
    \rho_\mathrm{o}^2)}{4 \delta}\,.
\end{equation}
Given these two parameters, the Machta-Zwanzig approximation is here again
given by Eq. (\ref{DMZ}). Note that in the present model, $\DMZ$
depends linearly on $\delta$, contrary to the Lorentz gas discussed in the
previous section, for which the area of the cells varied with $\delta$.

\subsection{Single-step memory approximation}

Consider a persistent random walk with single-step memory on a square
lattice. We denote by $\pf$, $\pr$, $\pb$ and $\pl$, respectively, the
conditional probabilities that the particle proceeds in the same direction,
turns right, reverses direction, or turns left. The velocity 
auto-correlation is 
\begin{equation}
  \mean{\v_0 \cdot \v_k} = \frac{1}{2}\{
     [\pf - \pb - \ii (\pr - \pl)]^k  + 
     [\pf - \pb + \ii (\pr - \pl)]^k \}\,.
    \label{v0vn1SMAsq} 
\end{equation}
Plugging this into (\ref{dcoef}) and assuming a symmetric
process such that $\pr = \pl \equiv \ps = (1 - \pf - \pb)/2$, we obtain the
diffusion coefficient 
\begin{equation}
 \Done = \DMZ\frac{1 + \pf - \pb} {1 - \pf + \pb}\,.
    \label{D1SMAsq}
\end{equation}
This expression compares to 
\begin{equation}
  \DoneKK = 1 + 2\mean{\v_0 \cdot \v_1} = \DMZ (1 + 2 \pf - 2 \pb)\,,
  \label{DKK1sq}
\end{equation}
which is the first-order approximation of Eq.~(\ref{D1SMAsq})
when expanding $\Done$ about the isotropic process for which $\pf = \pb =
1/4$.

\subsection{Two-step memory approximation}

Applying the same procedure to the persistent random walk with two-step
memory, we have $16$ conditional transition probabilities, in terms of which
it is possible to write the velocity auto-correlation in the compact form
\cite{GS09}
\begin{widetext}
  \begin{eqnarray}
    \mean{\v_0 \cdot \v_k} 
    = 2\Big(1\enspace 1\enspace 1\enspace 1\Big)
    &&\left[
      \left(
        \begin{array}{c@{\enspace}c@{\enspace}c@{\enspace}c}
          \pff & \plf & \pbf & \prf \\
          \ii \pfl & \ii \pll & \ii \pbl & \ii \prl \\
          - \pfb & - \plb & - \pbb & - \prb \\
          -\ii \pfr & -\ii \plr & -\ii \pbr & -\ii \prr 
        \end{array}
      \right)^{k-1}
      \left(
        \begin{array}{c@{\enspace}c@{\enspace}c@{\enspace}c}
          1 & 0 & 0 & 0 \\
          0 & \ii & 0 & 0 \\
          0 & 0 & -1 & 0 \\
          0 & 0 & 0 & -\ii
        \end{array}
      \right)
    \right.
    \nonumber\\
    &&+\left.
      \left(
        \begin{array}{c@{\enspace}c@{\enspace}c@{\enspace}c}
          \pff & \plf & \pbf & \prf \\
          -\ii \pfl & -\ii \pll & -\ii \pbl & -\ii \prl \\
          - \pfb & - \plb & - \pbb & - \prb \\
          \ii \pfr & \ii \plr & \ii \pbr & \ii \prr 
        \end{array}
      \right)^{k-1}
      \left(
        \begin{array}{c@{\enspace}c@{\enspace}c@{\enspace}c}
          1 & 0 & 0 & 0 \\
          0 & -\ii & 0 & 0 \\
          0 & 0 & -1 & 0 \\
          0 & 0 & 0 & \ii
        \end{array}
      \right)
  \right]
  \left(
    \begin{array}{c}
      p_1\\
      p_2\\
      p_3\\
      p_4
    \end{array}
  \right),
  \label{v0vk2SMAsq}
 \end{eqnarray}
where the $p_i$'s are the first four components of the stationary distribution
of
this process, which, assuming a left-right symmetric process, have the
expressions 
\begin{equation}
  \begin{array}{l}
    p_1 = \frac{\displaystyle \pbf \psb + \psf - \pbb \psf}
    {\displaystyle 4 (1 - \pff + \psb - \pff \psb + \pbf (-\pfb + \psb) + 
      \pbb (-1 + \pff - \psf) + \psf + \pfb \psf)},\\
    \\
    p_2 = \frac{\displaystyle 1 - \pbb - \pbf \pfb - \pff + \pbb \pff}
    {\displaystyle 8 (1 - \pff + \psb - \pff \psb + \pbf (-\pfb + \psb) + 
      \pbb (-1 + \pff - \psf) + \psf + \pfb \psf)},\\
    \\
    p_3 = \frac{\displaystyle \psb - \pff \psb + \pfb \psf}
    {\displaystyle 4 (1 - \pff + \psb - \pff \psb + \pbf (-\pfb + \psb) + 
      \pbb (-1 + \pff - \psf) + \psf + \pfb \psf)},\\
    \\
    p_4 = \frac{\displaystyle 1 - \pbb - \pbf \pfb - \pff + \pbb \pff}
    {\displaystyle 8 (1 - \pff + \psb - \pff \psb + \pbf (-\pfb + \psb) + 
      \pbb (-1 + \pff - \psf) + \psf + \pfb \psf)}.
  \end{array}
  \label{invpsq}
\end{equation}
\end{widetext}

The diffusion coefficient of this process may then be obtained by substituting
Eqs.~(\ref{v0vk2SMAsq})-(\ref{invpsq}) into Eq.~(\ref{dcoef}) and summing
over $k$. We will however not write down its lengthy explicit expression,
referring the reader instead to \cite{GS09}.

Let us also notice that Eqs.~(\ref{v0vk2SMAsq})-(\ref{invpsq}) can be used
to write down the corresponding second-order approximation in \cite{KK02},
here properly taking into account the stationary distribution,
\begin{equation}
  \DtwoKK = 1 + 2 \mean{\v_0\cdot\v_1} + 2 \mean{\v_0\cdot\v_2} 
    = \DoneKK + 2 \mean{\v_0\cdot\v_2}\,.  \label{DKK2sq}
\end{equation}

\subsection{Numerical results}

\begin{figure}[htb]
  \centering
  \includegraphics[width=0.45\textwidth]{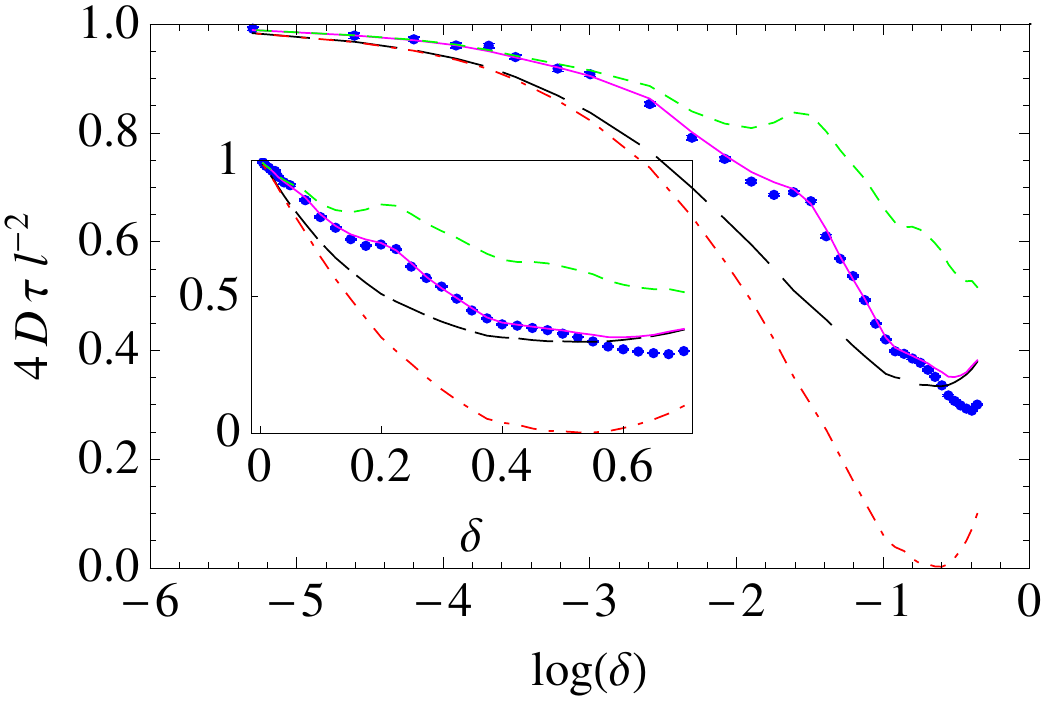}
  \caption{Numerical computation of the diffusion coefficient of the
    billiard table on a square lattice as shown in Fig.~\ref{fig.table-sq},
    here divided by the dimensional factor, Eq.~(\ref{DMZ}). The lines
    correspond to the different approximate results discussed above:
    (long-dashed, black line) single-step memory     approximation
    (\ref{D1SMAsq}); (dot-dashed, red line) first-order truncation 
    (\ref{DKK1sq}); (solid, magenta line) two-step memory  approximation,
    obtained from Eqs.~(\ref{dcoef}) and (\ref{v0vk2SMAsq}); (dashed, green
    line) second-order truncation (\ref{DKK2sq}).}  
  \label{fig.dcoeff-sq}
\end{figure}

The transition probabilities of the random walks with single- and two-step
memory approximation can be computed numerically for the billiard table
by estimating the relative frequencies of the corresponding events and
taking into account the left-right symmetry of these transitions. as above.
Plugging their values into
Eqs.~(\ref{D1SMAsq}) and (\ref{v0vk2SMAsq}), we obtain the corresponding
coefficients and compare them to the numerically-computed diffusion
coefficient of the billiard. These results are reported in
Fig.~\ref{fig.dcoeff-sq}, including the results of the truncations
(\ref{DKK1sq}) and (\ref{DKK2sq}). 

In the limit $\delta\to0$, we again see
that the Machta-Zwanzig approximation (\ref{DMZ}) is recovered. Zooming
into the lower range of the parameter $\delta$, we see that the single-step
approximation yields a good 
estimate of the actual diffusion coefficient of the Lorentz gas only for
values of $\delta \lesssim 2\times 10^{-3}$. The extent of the range of 
validity of the two-step approximation, on the other hand, is again much
larger, $\delta \lesssim 0.5$.

\begin{figure}[htb]
  \centering
  \includegraphics[width=0.45\textwidth]{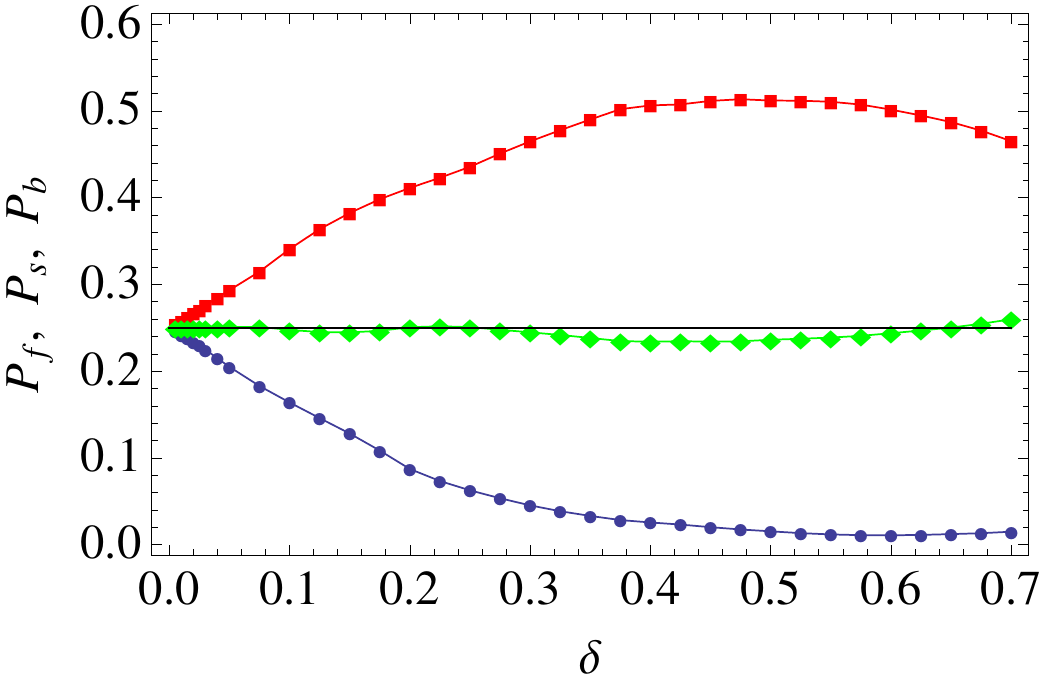}
  \caption{(Color online) Numerical computation of the transition
    probabilities $\pb$ (squares), $\pf$ (circles) and  $\ps$ (diamonds) of
    the single-step memory approximation associated to the billiard table on a
    square lattice. These rates reflect the anisotropy of the process.}
  \label{fig.corjump-sq}
\end{figure}

The other main observation is that the truncated estimates (\ref{DKK1sq})
and (\ref{DKK2sq}) are inaccurate as soon as $\delta \gtrsim 0.05$. The
reason can be traced to the anisotropy of the hopping processes.
Figure~\ref{fig.corjump-sq} shows the transition rates of the single-step memory
random walk. Although the probability of a right or left turn remains close
to $1/4$ throughout the parameter range, the backscattering probability
starts growing linearly above $1/4$ with small $\delta$'s and saturates
near $1/2$ at around $\delta = 0.5$. Correspondingly, the forward-scattering
probability decreases and is close to zero at around $\delta = 0.5$.

\begin{figure*}[hbt]
  \centering
  \includegraphics[width=0.3\textwidth]{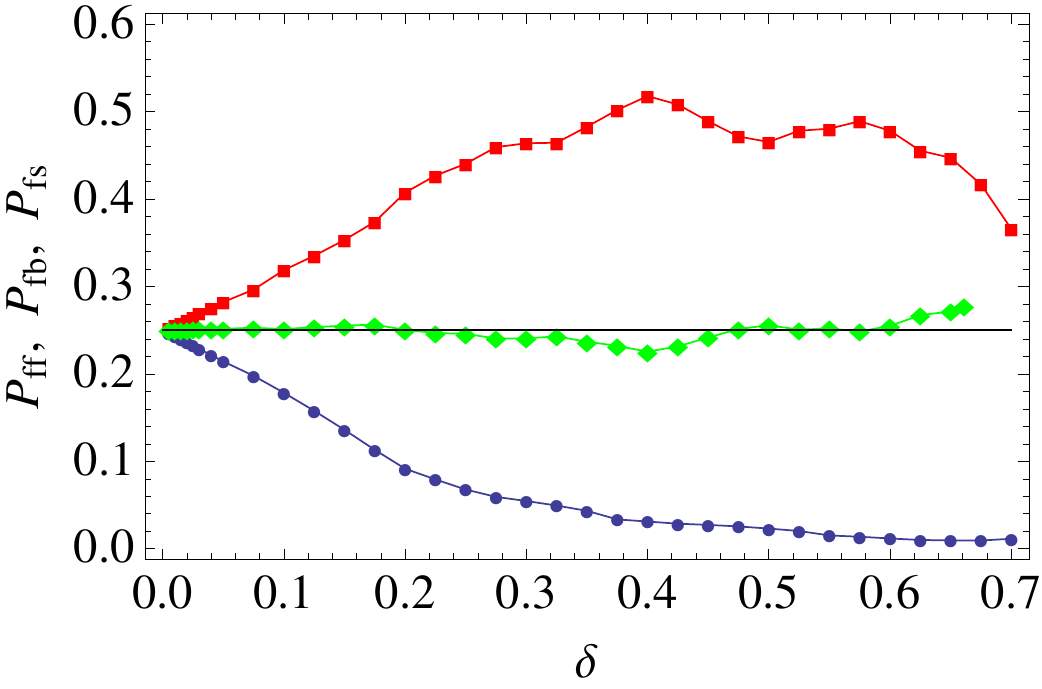}
  \includegraphics[width=0.3\textwidth]{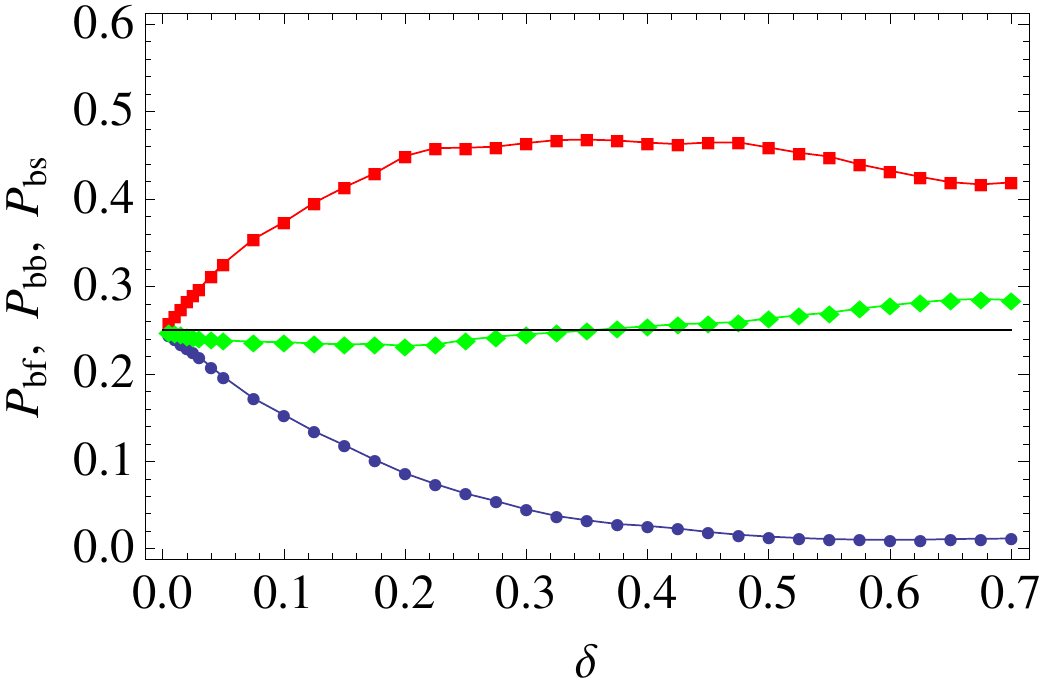}
  \includegraphics[width=0.3\textwidth]{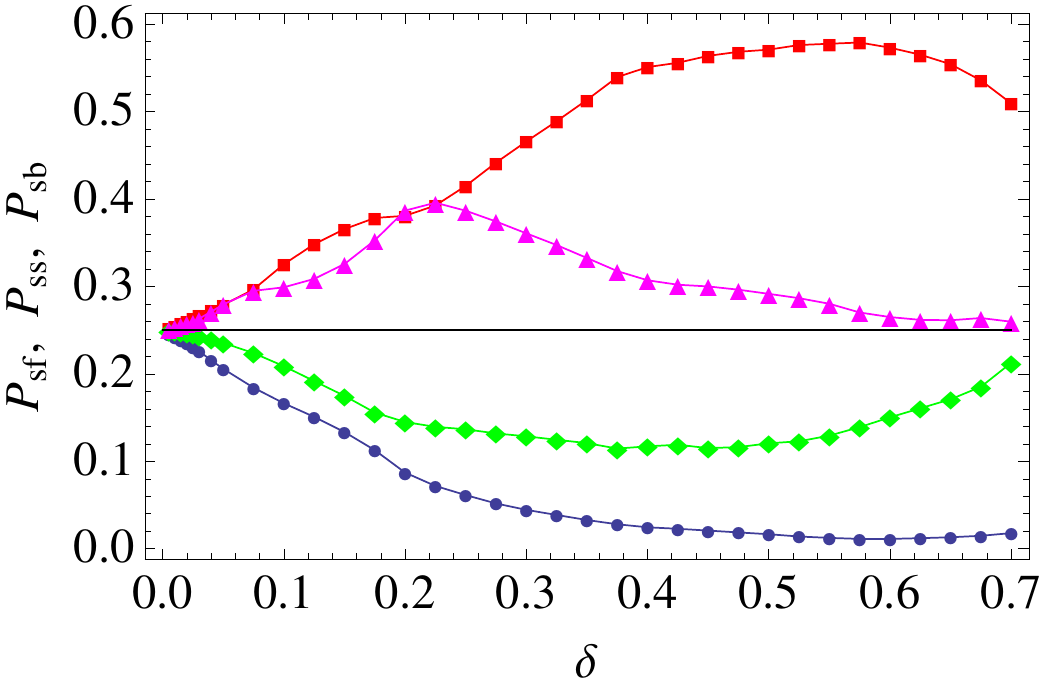}
  \caption{(Color online) Numerical computation of the transition
    probabilities $P_\mathrm{xb}$ (squares), $P_\mathrm{xf}$ (circles) and
    $P_\mathrm{xs}$ (diamonds) of the two-step memory approximation
    associated to the billiard table on a square lattice, where x stands
    respectively for f (left), b (middle) and s (right). In the right
    figure, the triangles are the probabilities of turning right after
    turning left and, the other way around, turning left after turning
    right. The differences between these three figures and the single-step
    transition probabilities shown in Fig.~\ref{fig.corjump-sq} justify
    resorting to a two-step memory process.}
  \label{fig.corjump2SMA-sq}
\end{figure*}

Looking at the transition probabilities of the two-step memory
process, shown in Fig.~\ref{fig.corjump2SMA-sq}, we notice the differences among
these probabilities, for instance
comparing $\pbb$, $\pfb$ and $\psb$, as well as between these probabilities
and that of the single-step memory process, in our example $\pb$. These
differences justify the necessity of resorting to the two-step memory
process over the
single-step process.

\section{Conclusions}

The diffusion coefficients of deterministic systems with rapid decay of
correlations can be well approximated by that of correlated walks, where
a walker's transition probabilities are determined according to its motion
over the last few steps.

Billiards provide good, physically motivated, examples of such systems. The
Machta-Zwanzig dimensional prediction \cite{MZ83}, according to which the
diffusion coefficient is approximated by the ratio between the distance
between traps squared and the trapping time, provides a gross estimate of
this quantity. However, the importance of memory effects in the
deterministic diffusion of tracer particles is apparent as soon as one
moves away from the limit where the trapping times are infinite.

Truncation schemes based on the Green-Kubo formula, such as considered in
\cite{Kla07,KK02}, may provide accurate results for models with little
anisotropy, but they are physically inconsistent: given a hopping process
with finite memory effects, velocity auto-correlations of all orders yield
non-vanishing contributions to the Green-Kubo formula.

This is particularly clear where anisotropies come into play. Estimates of
the diffusion coefficients based on persistent random walks, however, do
provide accurate results where the truncation schemes breakdown.

It will be interesting to find out how these results transpose to models
where disorder is present, such as with tagged-particle diffusion in
interacting particle systems. Dimensional predictions similar to the
Machta-Zwanzig one also appeared recently in the context of models of heat
conduction \cite{GG08, GL08}. Estimating the deviations of the heat
conductivities of these models from dimensional predictions remains an open
problem.

\begin{acknowledgments}
The authors thank Felipe Barra, Mark Demers, Hern\'an Larralde and
Carlangelo Liverani for helpful discussions. This research benefitted from
the joint support of FNRS (Belgium) and CONACYT (Mexico) through a
bilateral collaboration project. The work of TG is financially supported by the
Belgian Federal
Government under the Inter-university Attraction Pole project NOSY
P06/02. TG is financially supported by the Fonds de la Recherche
Scientifique F.R.S.-FNRS.  DPS acknowledges financial support from
DGAPA-UNAM project  IN105209, and the hospitality of the Universit\'e Libre de
Bruxelles,
where 
most of this work was carried out.  TG acknowledges
the hospitality of the Weizmann Institute of Science, where part of this work
was completed.
\end{acknowledgments}

\end{document}